\documentclass[aps,twocolumn]{revtex4}

\begin{document}

\def\bb{\begin{equation}}
\def\ee{\end{equation}}

\title{Spin-fluctuations in the quarter-filled Hubbard ring
: significances to LiV$_2$O$_4$}
\author{J.D. Lee}
\address{W.M. Keck Laboratories, California Institute of Technology,
Pasadena, CA 91125}
\date{\today}

\begin{abstract}

Using the quantum Monte Carlo method, we investigate 
the spin dynamics of itinerant electrons in the one-dimensional
Hubbard system. Based on the model calculation, we have studied
the spin-fluctuations in the quarter-filled metallic Hubbard ring,
which is aimed at the vanadium ring or chain defined along corner-sharing 
tetrahedra of LiV$_2$O$_4$, and found the dramatic changes of 
magnetic responses and spin-fluctuation characteristics with the temperature.
Such results can explain the central findings in the recent neutron
scattering experiment for LiV$_2$O$_4$.

\end{abstract}


\maketitle

LiV$_2$O$_4$ is a transition metal oxide with a cubic spinel 
(or pyrochlore) structure
showing many essential features of the heavy-fermion system like
Ce compounds\cite{Kondo}. Its specific heat coefficient is the largest
one observed among other 3$d$ metallic systems, 
$\gamma\sim 420{\rm mJ/mol\hbox{ }K}^2$.
It has become an important issue to clarify the physical origin
of a high density of low-energy fermionic excitations
without localized $f$-levels.  There's also a great interest in the
unusual magnetic properties of LiV$_2$O$_4$ due to the itinerant
frustrated nature. 
Another metallic system, Y(Sc)Mn$_2$\cite{Shiga},
has also been focused on as a frustrated spin liquid belonging to a similar
class to LiV$_2$O$_4$. Y(Sc)Mn$_2$ exhibits many similarities;
it is the geometrically frustrated magnet with no long range order, 
nearly antiferromagnetic itinerant system, 
and most interestingly the heavy fermion system.

Three classes of theoretical mechanisms are most
frequently referred to understand the heavy-fermion properties
in LiV$_2$O$_4$. One is the spin-fluctuations in the three-dimensional
frustrated lattice as in the study of Y(Sc)Mn$_2$\cite{Canals}.
Due to the magnetic frustrations, the spin cannot order down to low
temperatures, resulting in the large fermionic entropy.
The other is the well-known Kondo effect.
From the band structure calculations\cite{Matsuno,Anisimov},
it is shown that $3d$ $t_{2g}$ bands of V are crossing the Fermi level
and, by the trigonal crystal field, split to a bit narrow-band
half-filled $A_{1g}$ singlet and a bit wide-band quarter-filled 
$E_{g}$ doublet. These results
lead to the mapping of the electronic structure into the Kondo
lattice model\cite{Anisimov}. 
The third candidate is the mechanism based
on the one-dimensional electronic structure, where it is expected 
that the correlation effect is much enhanced, giving 
the large specific-heat coefficient. Fulde {\it et al.}\cite{Fulde}
have suggested that the large $\gamma$ coefficient results from
excitations of Heisenberg spin 1/2 chains and rings, 
which are by the direct consequence of the frustration of
corner-sharing tetrahedra of the vanadium lattice.
Their idea on the formation of spin chain or ring
in LiV$_2$O$_4$ dates back to the study on Yb$_4$As$_3$\cite{Fulde95},
where, due to a charge ordering of Yb ions,
the electronic structure could be interpreted as well-decoupled
(at least magnetically) one-dimensional chains.

In the last decades, the spin-fluctuation has been found to
play fundamental roles in many of strongly correlated electron systems,
especially in the high $T_C$ superconductors\cite{sf}.
The combined system of the strong electron correlation and
the spin itineracy, where there is no magnetic long range order,
leads to intriguing spin-fluctuations.
Recent two inelastic 
neutron scattering experiments\cite{Krimmel,Lee} have delivered
seminal informations on the spin dynamics in LiV$_2$O$_4$;
(i) they have reported the dramatic crossover from a ferromagnetic (FM)
to an antiferromagnetic (AFM) spin-fluctuation with the temperature $T$
lowered.
Especially, Lee {\it et al.}\cite{Lee} have explicitly pointed out
the AFM spin-fluctuation would be centered around $Q_c=0.64\AA^{-1}
=0.59\pi/a$ ($a$ is the V-V distance), (ii) they have 
found the residual relaxation rate for $T\to 0$ and its monotonous
increase at $Q_c$ with raising temperature, but the increasing behavior
was reported differently from each other, i.e.
Krimmel {\it et al.}\cite{Krimmel} have reported the square-root
temperature behavior, whereas Lee {\it et al.}\cite{Lee} the linear behavior,
and (iii) Krimmel {\it et al.} have also provided 
the momentum-transfer-dependence of the relaxation rate to elucidate
the change of spin-fluctuation characteristics.

In this paper, we bring focus on the spin dynamics of LiV$_2$O$_4$
grounded on the one-dimensional mechanism.
It is actually clear that the Heisenberg spin 1/2 chain by itself cannot
explain the observed experiments because the low energy excitation of
the system should be well-dispersive gapless magnon.
Instead, the quarter-filled Hubbard ring is taken as the starting point,
which gives the itineracy to Fulde's spins,
toward an understanding of characteristic spin-fluctuations
observed in LiV$_2$O$_4$.
Fujimoto\cite{Fujimoto} has studied the network of quarter-filled 
Hubbard chains accounting for the hybridization between chains
along the similar line to the Fulde's spin chain or ring.
In the study, he has obtained quite a comparable size of $\gamma$ 
to the experimental value and introduced another 
energy scale $T^{\ast}$ giving the dimensional crossover; 
below $T^{\ast}$, three-dimensional Fermi-liquid state with the heavy
mass is realized, but above $T^{\ast}$, one-dimensional characters 
dominate. He has also pointed out that Urano {\it et al.}\cite{Urano}'s 
transport data can be consistent with this picture and a characteristic
low temperature scale ($\sim 20$ K) observed would correspond 
to $T^{\ast}$. In the present investigation, therefore, 
we do not consider the very low temperature region 
of $T\ll T^{\ast}$\cite{TK}.

In an actual situation of LiV$_2$O$_4$, the one-dimensional ring (chain) is
constructed on the corner-sharing tetrahedra of the vanadium network,
which has been visualized in Refs.\cite{Fulde,Fujimoto}.
The Hubbard model is defined on the one-dimensional lattice;
\begin{eqnarray}\label{model}
{\cal H}&=&-t\sum_{i}\sum_{\delta=\pm 1}\sum_{\sigma}
         (c_{i\sigma}^{\dagger}c_{i+\delta\sigma}+{\rm H.c.})
         -\mu\sum_i(n_{i\uparrow}+n_{i\downarrow})
\nonumber \\
        &+&U\sum_in_{i\uparrow}n_{i\downarrow},
\end{eqnarray}
where $c_{i\sigma}^{\dagger}$ and $c_{i\sigma}$ are the creation and
annihilation operators for electrons with spin $\sigma$ at lattice $i$
and $n_{i\sigma}=c_{i\sigma}^{\dagger}c_{i\sigma}$.
$t$ is the hopping parameter, $\mu$ the chemical potential, and
$U$ the on-site Coulomb correlation. 
The first and the last sites are connected by imposing the periodic
boundary condition, i.e. it gives the "ring" geometry.

Among many versions of quantum Monte Carlo (QMC) methods, 
it is the path integral theory of QMC
that is better proper for a description of the itinerant
electron systems\cite{Blank,White_a,Tomita}. The path integral QMC
for the correlated electron system cooperates the Hubbard-Stratonovitch
transformation and integrates out the electron field.
Most of all the QMC methods are based on the Trotter decomposition
$e^{-\beta(K+V)}\approx (e^{-\Delta\tau V}e^{-\Delta\tau K})^{L}$
with $\beta=1/T=\Delta\tau L$\cite{Suzuki}
where $K$ should be the one-electron terms and $V$
the electron-electron correlation term.
The collective spin excitations probed by the inelastic neutron scattering
are described in the time-dependent spin-spin correlation function
$S({\bf q},\tau)$
\bb\label{Sqtau}
S({\bf q},\tau)=\frac{1}{N}\sum_{ij}e^{i{\bf q}\cdot({\bf R}_i-{\bf R}_j)}
\langle [n_{i\uparrow}(\tau)-n_{i\downarrow}(\tau)]
        [n_{j\uparrow}-n_{j\downarrow}]\rangle,
\ee
which is corresponding to the thermodynamic two-particle Green's function
in the imaginary time. Through the analytic continuation from
$S({\bf q},\tau)$ under a condition $S({\bf q},\omega)\ge 0$\cite{White_b},
we obtain the experimentally observable spectral function
$S({\bf q},\omega)$ satisfying
\bb\label{cont}
S({\bf q},\tau)=-\int\frac{d\omega}{2\pi}\frac{e^{-\omega\tau}}
{1-e^{-\beta\omega}}S({\bf q},\omega).
\ee

For a numerical simulation in the study,
the 24-site Hubbard ring with $U/t=4$ is considered.
We follow the basic approach
for the grand canonical ensemble. For the quarter-filled occupation,
the ensembles such that it can give $1/N\sum_i(n_{i\uparrow}
+n_{i\downarrow})=0.5$ should be sampled by adjusting the value of $\mu$.
The Trotter decomposition is done such that $d\tau=0.1$
(only for a case of $1/T=1.5$, $d\tau=0.05$).
We have taken the averages of the dynamical correlation functions
over $10^4$ updates of all the Hubbard-Stratonovitch bosons on the lattice.
Further, to keep the numerical stability, we have used
the matrix factorization technique\cite{White_a}.
All the energy quantities are measured in a unit of $t$ and
all the momentum quantities in $\pi/a$.

\begin{figure}
\vspace*{7.5cm}
\includegraphics{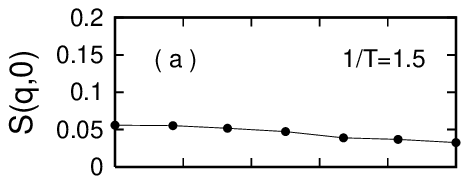}
\caption{(a)-(e) Dynamical spin-spin correlation function
in the quarter-filled Hubbard ring as a function of $q$
in the quasi-elastic mode ($\omega=0$) at several temperatures,
i.e. $S(q,\omega=0)$. (f) Ratio of $S(q=1/2,\omega=0)$ and
$S(q=0,\omega=0)$.
}
\label{FIG1}
\end{figure}

Figure \ref{FIG1}(a)-(e) 
show the dynamical spin-spin correlation function 
in the quasi-elastic mode of the quarter-filled Hubbard ring at various 
temperatures. As the temperature decreases, at first, the AFM 
short-range order, induced by the AFM spin-fluctuation, with  
a characteristic wave vector $q=1/2$ becomes appreciable.
On the other hand, in the high temperature region, its character
is found rather FM, more obvious  
in Fig.\ref{FIG3}. This finding is very consistent with
the dramatic crossover of the character of spin-fluctuation 
with the temperature reported by recent inelastic neutron 
scattering experiments\cite{Krimmel,Lee}. Further, the characteristic 
wave vector $q=1/2$ is also moderately consistent with 
the staggered wave vector $Q_c=0.59$ found in the experiment\cite{Lee}.
It can be understood as follows; 
the one-dimensional spin-spin correlation function 
in the itinerant limit gives rise to a logarithmic singularity
at $2k_F$ ($k_F$ is the Fermi wave vector;
$k_F=1/4$ in the present case)\cite{Ogata}, from which 
in the correlated limit
incomplete magnetic moments (i.e. AFM spin-fluctuation) 
could evolve at every other site and lead to the AFM short-range order.
In the low temperature region, $S(q,0)$ is rapidly suppressed
for high momentum transfers ($q>1/2$). In Fig.\ref{FIG1}(f),
the ratio of quasi-elastic peaks of the magnetic scattering
at $q=0$ and $q=1/2$ shows roughly the change of characteristic
spin-fluctuations governing the system around $1/T\sim 4$ or
$T\sim 0.25$. Here we need estimate the size of $t$. 
Fulde {\it et al.}\cite{Fulde} 
have estimated the exchange integral of $J$ in the Heisenberg spin 1/2 chain
on the frustrated lattice as about 3 meV\cite{Fulde}. Noting
$J\sim {\cal O}(t^2/U)$, we then have approximately $t\sim 10$ meV.
Now the crossover temperature is $T\sim 0.25\sim 30$ K, close
to Krimmel {\it et al.}'s 40 K\cite{Krimmel}. 

\begin{figure}
\vspace*{4.9cm}
\includegraphics{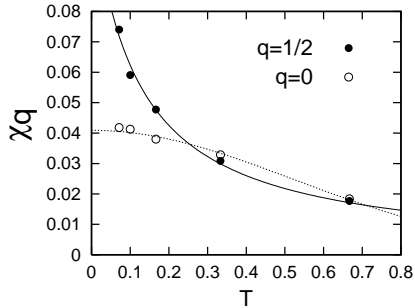}
\caption{Crossover from FM to AFM spin-fluctuation with $T$.
The static spin susceptibility at $q=0$ and $q=1/2$;
the solid line for $q=1/2$ is the AFM Curie-Weiss susceptibility
and the dotted line for $q=0$ is a guide for the eye.
}
\label{FIG2}
\end{figure}

To scrutinize the changes of magnetic responses more,
we provide, in Fig.\ref{FIG2}, the static susceptibility $\chi_q$ at $q=0$
and $q=1/2$ with $T$. The static susceptibility
$\chi_{\bf q}$ can be directly evaluated from $S({\bf q},\tau)$
obtained by the QMC calculation
\bb\label{chi}
\chi_{\bf q}=\int_0^{\beta}d\tau S({\bf q},\tau).
\ee
It is shown in the figure that $\chi_q$ for $q=1/2$ nicely 
follows the AFM Curie-Weiss susceptibility, $\propto 1/(T+\theta)$.
Such Curie-Weiss behavior has been already observed in the
experiment\cite{Lee}, where $\theta$ for the best fitting was estimated
as 7.5 K. Our calculation gives the similar value, 
$\theta\sim0.11\sim 13.2$ K.
Let us note it means that the quarter-filled one-dimensional Hubbard
model could allow a formation of magnetic moments like Kondo lattice model
in the high temperature\cite{Hewson}, where the localized level manifests
the Curie-Weiss susceptibility. {\it Unstable} magnetic moments
produce the AFM short-range order
centered around $\pi/(2a)$ (i.e. $q=1/2$), where $2a$ is the distance
between neighboring moments. It is consistent with the
temperature behavior of the magnetic relaxation rate 
$\Gamma_q$ at $q=1/2$ in Fig.\ref{FIG3}.
Fig.\ref{FIG2} also shows the crossover of magnetic characters from FM to AFM
around $T\sim 0.25\sim 30$ K, consistent with Fig.\ref{FIG1}(f).

\begin{figure}
\vspace*{8.5cm}
\includegraphics{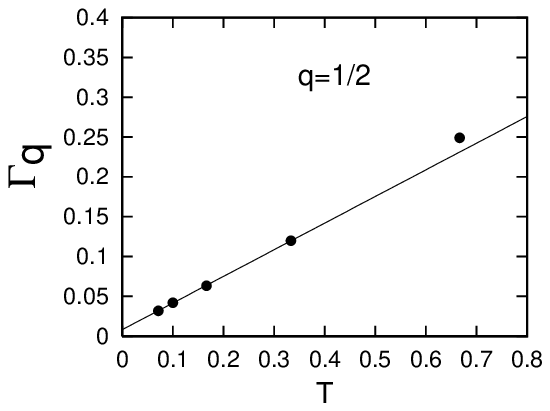}
\caption{Magnetic relaxation rate $\Gamma_{\bf q}$.
Upper panel: the temperature dependence of $\Gamma_{\bf q}$ at $q=1/2$.
Lower panel: the $q$-dependence of $\Gamma_{\bf q}$ at high and low 
temperatures.
}
\label{FIG3}
\end{figure}

Another important quantity is the magnetic relaxation rate
$\Gamma_{\bf q}$ usually defined by the simple ansatz for the
dynamic susceptibility
\bb\label{Sq_ansatz}
S({\bf q},\omega)=\frac{1}{1-e^{-\omega/T}}
\frac{\omega\Gamma_{\bf q}\chi_{\bf q}}{\omega^2+\Gamma_{\bf q}^2}.
\ee
In the study, $\Gamma_{\bf q}$ is evaluated by taking $\omega\to 0$ 
in Eq.(\ref{Sq_ansatz}), 
$$
S({\bf q},0)=\frac{T\chi_{\bf q}}{\Gamma_{\bf q}},
$$
where $S(q,0)$ and $\chi_q$ are already given in
Figs.\ref{FIG1} and \ref{FIG2}, respectively. But it should be noted
that, because the unit of $S(q,0)$ is arbitrary\cite{unitofS},
$\Gamma_q$ would be obtained only up to a constant.
That is, we note the true relaxation rate should be $\eta\Gamma_q$
($\eta$ is a nonzero constant).
The results of $\Gamma_q$ are provided in Fig.\ref{FIG3}.
The upper panel of Fig.\ref{FIG3} shows that the spin magnetic
relaxation rate $\Gamma_q$ at $q=1/2$ increases linearly with temperature
for a rather wide temperature range to $\sim 0.7\sim 80$ K.
The linear increasing behavior for such a wide $T$ 
range (to $\sim 80$ K) has been 
ascertained in the experiment by Lee {\it et al.}\cite{Lee}, where
its increasing rate is found 0.46. 
Linear $T$ behavior is actually unusual in the $f$-electron heavy fermion
system. It is however a bit common in frustrated metal oxides,
normally related with {\it unstable} local moments.
In the Kondo system, one most usually has $\Gamma_q\sim \Gamma_q^0+bT^{1/2}$, 
where $\Gamma_q^0\sim T_K$ (Kondo temperature)\cite{Hewson}.
Earlier, through mapping into the Kondo model, 
Anisimov {\it et al.}\cite{Anisimov} have estimated
$T_K\sim 550$ K for the single-site case,
but argued that another characteristic energy scale
$T_{coh}(\sim 25-40$ K, comparable to $\Gamma_q^0$) 
would replace $T_K$ in the dense Kondo lattice.
The increasing rate of the present result shown in Fig.\ref{FIG3}
is estimated as 0.32 and $\Gamma_q^0$ at $T=0$ is obtained about 0.1 meV.
Comparing the increasing rates, we find an unknown constant $\eta$
should be about 1.44 and the residual relaxation rate 
$\Gamma_q^0$ be 0.14 meV.
This value is smaller than experimental findings, i.e. 
Krimmel {\it et al.}\cite{Krimmel} have reported 0.5 meV
and Lee {\it et al.}\cite{Lee} 1.4 meV. 
The lower panel of Fig.\ref{FIG3} gives the $q$ dependences of $\Gamma_q$,
whose dependences also agree with
an observation of the experiment\cite{Krimmel}.
For $1/T=1.5$ (high temperature), $\Gamma_q$
shows a linear $q$ dependence, which is actually expected 
in the spin-fluctuation theories of weak FM metals\cite{Moriya}.
Therefore, the behavior of linear $q$ dependence 
is consistent with our argument that the system
should be a metal with weak FM spin-fluctuations at 
high temperatures ($\gtrsim 0.25$), being 
associated with Figs.\ref{FIG1} and \ref{FIG2}. On the other hand,
$\Gamma_q$ at the low temperature ($1/T=10$) is almost constant
for small $q$'s ($q\le 1/2$), but rapidly increases for high $q$'s 
($q>1/2$). The constant $\Gamma_q$ with $q$ at low $T$ 
(not anticipated by the simple Fermi-liquid theory)
was found in the experiment, but a rapid increase for high $q$'s
was not, which instead may be attributed to another subtle feature 
of one-dimensional Hubbard model. Recently, it has been found that
the nonlinear coupling between spin and charge in the one-dimensional
Hubbard model would lead to coupled collective excitations other than 
spin-fluctuations (or magnons) in $S(q,\omega)$
especially for high $q$'s\cite{Tomita}. Those may serve 
as additional decaying channels for spin fluctuations.
It is noted that such enhanced $\Gamma_q$ is directly connected
with the diminution of $S(q,0)$ at low $T$ for $q>1/2$
in Fig.\ref{FIG1}.

In summary, we have discussed the recent inelastic neutron
scattering experiments for LiV$_2$O$_4$ based on the QMC study
of spin-fluctuations in the quarter-filled Hubbard ring.
In the study, neutron scattering cross sections, static spin 
susceptibilities, and spin relaxation rates have been evaluated
with temperatures and momentum transfers. They are
found quite consistent with the experiment qualitatively,
or semi-quantitatively. 
Particularly, it is appealing that the AFM short-range correlation
develops due to a formation of unstable magnetic moments 
as $T$ decreases in the quarter-filled Hubbard ring, 
which explains the AFM spin-fluctuation
around $Q_c=0.59$ observed in LiV$_2$O$_4$. Our finding that
a single quarter-filled Hubbard ring can explain the neutron scattering
experiment could be consistent with a case of
decoupled chains in Yb$_4$As$_3$ 
unless we think of the very low temperature regime.
However, it is a difference that the one-dimensional electronic 
structure is expected from the geometrical frustration
in LiV$_2$O$_4$\cite{Fujimoto}.
We would like to remark that at least a few
experimental findings cannot be explained by the Kondo lattice model,
but by the one-dimensional quarter-filled metallic model; 
(i) features of nearly-FM metal at high temperatures,
(ii) the linear $T$ dependence of $\Gamma_q$ for a wide $T$ range, 
and (iii) the constant $\Gamma_q$ with $q$ at low temperatures
(actually decreasing behaviors in CeCu$_6$\cite{Aeppli}).
Further, the recent nuclear magnetic resonance (NMR)
study for LiV$_2$O$_4$ under high pressure
has reported an opposite behavior of $T_1$ (spin-lattice relaxation
time) to that of Ce compounds\cite{Fujiwara}.
It is also worth stressing that magnetic responses (such as 
$T$ dependence of $\chi$) of LiV$_2$O$_4$ differ qualitatively
from Y(Sc)Mn$_2$\cite{Nakamura}.
Therefore, the present results can be one of evidences 
along with other studies\cite{Fulde,Fujimoto} that 
LiV$_2$O$_4$ comprises one-dimensional chains or rings and behaves
like the one-dimensional system. Finally, it should be noted that
the conclusion casts another important problem to us. 
It is well known that, in one-dimensional metallic system,
some extreme realization of correlation effects like
the spin-charge separation occurs, which has been actually observed 
in SrCuO$_2$ by the photoemission spectroscopy\cite{Kim}.
Thus it must be fascinating to search for
the spin-charge separation in LiV$_2$O$_4$ by the photoemission
spectroscopy.
\\
- The author is indebted to Norikazu Tomita
for his enthusiastic helps on the technical side of QMC methods.
Discussions with Atsushi Fujimori, Brent Fultz, 
Seung-Hun Lee, ByungIl Min are greatly appreciated.
This work was supported by the U.S. Department of Energy under grant
number DE-FG03-01ER45950.


\begin{references}

\bibitem{Kondo} S. Kondo {\it et al.}, Phys. Rev. Lett. {\bf 78}, 3729 (1997); 
                S. Kondo {\it et al.}, Phys. Rev. B {\bf 59}, 3729 (1999).
\bibitem{Shiga} M. Shiga {\it et al.}, J. Phys. Soc. Jpn. 
                {\bf 62}, 1329 (1993); R. Ballou {\it et al.},
                Phys. Rev. Lett. {\bf 76}, 2125 (1997).
\bibitem{Canals} B. Canals and C. Lacroix, Phys. Rev. Lett. {\bf 80},
                 2933 (1998).
\bibitem{Matsuno} J. Matsuno {\it et al.}, Phys. Rev. B
                  {\bf 60}, 16359 (1999).
\bibitem{Anisimov} V.I. Anisimov {\it et al.}, Phys. Rev. Lett. {\bf 83},
                   364 (1999).
\bibitem{Fulde} P. Fulde {\it et al.}, Europhys. Lett. {\bf 54}, 779 (2001).
\bibitem{Fulde95} P. Fulde {\it et al.}, Europhys. Lett. {\bf 31}, 323 (1995).
\bibitem{sf} A.J. Millis {\it et al.}, Phys. Rev. B {\bf 42},
             167 (1990); J.D. Lee and A. Fujimori, Phys. Rev. Lett.
             {\bf 87}, 167008 (2001).
\bibitem{Krimmel} A. Krimmel {\it et al.},
                  Phys. Rev. Lett. {\bf 82}, 2919 (1999).
\bibitem{Lee} S.-H. Lee {\it et al.},
              Phys. Rev. Lett. {\bf 86}, 5554 (2001).
\bibitem{Fujimoto} S. Fujimoto, Phys. Rev. B {\bf 65}, 155108 (2002).
\bibitem{Urano} C. Urano {\it et al.}, Phys. Rev. Lett. {\bf 85}, 1052 (2000).
\bibitem{TK} $T^{\ast}$ could be ascribed to $T_K$ (Kondo temperature)
             within the scenario of Kondo lattice model.
\bibitem{Blank} R. Blankenbecler {\it et al.}, 
                Phys. Rev. D {\bf 24}, 2278 (1981); J.E. Hirsch, 
                Phys. Rev. B {\bf 31}, 4403 (1985).
\bibitem{White_a} S.R. White {\it et al.}, Phys. Rev. B {\bf 40},
                  506 (1989). 
\bibitem{Tomita} N. Tomita and K. Nasu, Phys. Rev. B {\bf 56}, 3779 (1997);
                 N. Tomita and K. Nasu, Phys. Rev. B {\bf 61}, 2488 (2000).
\bibitem{Suzuki} {\it Quantum Monte Carlo Methods}, edited by M. Suzuki
                 (Springer-Verlag, 1987).
\bibitem{White_b} S.R. White {\it et al.}, Phys. Rev. Lett. {\bf 63},
                1523 (1989); R.N. Silver {\it et al.},
                Phys. Rev. B {\bf 41}, 2380 (1990).
\bibitem{Ogata} M. Ogata and H. Shiba, Phys. Rev. B {\bf 41}, 2326 (1990).
\bibitem{Hewson} {\it The Kondo Problem to Heavy Fermions}, A.C. Hewson
                 (Cambridge University Press, 1993).
\bibitem{unitofS} The absolute value of $S({\bf q},\omega)$ is arbitrary.
                 The analytic continuation gives a series of delta functions
                 for $S({\bf q},\omega_i)$ at discrete $\omega_i$'s,
                 from which $S({\bf q},\omega_i)$ is convoluted
                 with a suitable profile function.
\bibitem{Moriya} {\it Spin fluctuations in Itinerant Electron Magnetisms},
                 T. Moriya (Springer-Verlag, 1989). 
\bibitem{Aeppli} G. Aeppli {\it et al.}, Phys. Rev. Lett.
                 {\bf 57}, 122 (1986).
\bibitem{Fujiwara} K. Fujiwara {\it et al.}, Physica B,
                   {\bf 312}-{\bf 313}, 913 (2002).
\bibitem{Nakamura} H. Nakamura {\it et al.}, J. Phys.: Condens. Matter,
                   {\bf 9}, 4701 (1997).
\bibitem{Kim} C. Kim {\it et al.}, Phys. Rev. Lett.
              {\bf 77}, 4054 (1996).



\end{references}
\end{document}